# Funding for Adaptive Optics in the United States by the National Science Foundation 2006-2009: An Update

by Jay A. Frogel

Association of Universities for Research in Astronomy

## Abstract

In 2006 I published an article in *GeminiFocus* that examined funding for astronomical adaptive optics related technology and instrumentation in the United States from 1995 through mid-2006. That article concluded that based on projections then current, AO implementation on public and private telescopes in the U.S. will soon seriously lag that on the ESO VLT as measured by funds available. It called for a significant infusion of public funds for AO development and implementation so that when combined with private funds, the U.S. astronomical community as a whole would be able to take full advantage of AO systems on both public and private telescopes. In 2006 I estimated that the total amount of public (NSF) funds that would be available in 2009 for AO related non-science activities would be about $6M.

This article updates the analysis done in my previous article. I show that for 2009 the funds for AO related non-science activities are about $7M in spite of the termination of the AODP program. Federal stimulus funds (ARRA) to the NSF and its grant programs account for a not insignificant part of this $7M. I make the probably optimistic prediction that in 2010 there will be just over $6M in NSF funds available for AO related non-science work.

Thus there has been no significant real increase of public funding for AO development and implementation since the predictions made in 2006. If private funding in the US and the level of ESO AO funding is close to the values predicted in my previous article, then *ESO on one observatory, will be outspending all US AO efforts spread over about a dozen observatories by a factor of three*.

## Introduction

In 2006 I published an article in *GeminiFocus*[1] (Frogel 2006, hereafter referred to as "F06") that examined funding for adaptive optics (AO) research and development (R&D), AO systems, and AO instrumentation in the United States from 1995 through mid-2006. This funding for *non-science* related AO activities is spread over more than one dozen telescopes and institutions in the US, but concentrated in just a few. In a typical year nearly 60% of these funds went to "private" observatories, i.e. those with limited or restricted access to their facilities. Observatories connected with public, state universities are considered private.

Over the years examined, typically 50 to 60% of funds for AO R&D and instrumentation in the US came from public sources – almost exclusively the NSF; the remaining 40% came from private sources such as foundations and educational institutions (Fig. 5 of F06). Typically, since 2001, the yearly amount spent on non-science AO[2] has varied between $10M and $20M from *all* sources (see Fig. 4 of F06). On average, about 25% of the NSF funds has been for AO activities at the Gemini Observatory. The other Gemini partner countries contribute an equal amount.

Based on several assumptions and extrapolations I predicted in F06 that from a high of about $10M per annum in 2004 to 2006, the yearly *public* funds available for AO would drop to about $6M per year by the end of 2009 and that public plus private expenditures would decline from a high of nearly $20M per annum to about $10M in 2009. By way of contrast, the European Southern Observatory, ESO, was expected to be spending between $30 and $40M per year on AO *on the VLT alone*.

Since we are now well into 2009 it is a good time to compare the predictions and extrapolations I made three years ago with what has actually happened. It is most straight-forward to do this for the public, i.e. NSF, funds expended since the relevant numbers are on the NSF or NOAO public web sites. This article present the update to F06.

## Important AO-related Activities in the US since 2006

Before presenting the update to F06, I will mention the main AO-related activities that have taken place in the US since 2006.

- In 2008 a new "Roadmap for the Development of United States Adaptive Optics" was made public[3]. This roadmap is based on a workshop held in 2007 with contributions from 24 participants and comments from more than 100 individuals in the astronomical community.

---

[1] Jay A. Frogel, "A History of Funding for AO in the United States", in *GeminiFocus*, December, 2006, pp 82-93. Also at http://www.aura-astronomy.org/nv/Astro2010PanelDocs/A%20History%20of%20Spending%20for%20AO%20in%20the%20US.pdf

[2] For the remainder of this article "AO funding" will imply only those funds used to support R&D and instrumentation for AO and will specifically *exclude* grants and other monies from all sources used to support *science* with AO systems on telescopes.

[3] Dekany, R. & Lloyd-Hart, M. editors, 2008, "A Roadmap for the Development of United States Astronomical Adaptive Optics", http://www.aura-astronomy.org/nv/AO_Roadmap2008_Final.pdf

- Also in 2008 AURA prepared a white paper on AO entitled "AURA's Assessment of Adaptive Optics: Present State and Future Prospects"[4]. This paper examined both AO technology and the science being done with AO systems. It is one of AURA's major pieces of input to Astro2010, the new Astronomy and Astrophysics Decadal Survey Committee.
- A number of brief white papers on various AO technology issues were also submitted to Astro2010 by AURA and others. These may be found at http://sites.nationalacademies.org/BPA/BPA_049492 and http://sites.nationalacademies.org/BPA/BPA_049522 .
- Adaptive optics figured prominently in some of the presentations made to the Astro2010 Panels[5].
- On the science front, a well attended "Meeting within a Meeting" was held at the 2009 June AAS meeting in Pasadena. Most of the more than 30 talks focused on recent science results obtained with AO instruments on Gemini, Keck, MMT, and the VLT.[6]

## **Public Funding For AO in the United States**:

By public funding I mean funds that are generally available to *all* astronomers via competitive, peer reviewed processes. These funds are for use at both public and private facilities. In the U.S. the NSF is the only significant *ultimate* source of *public* funds for AO development and implementation. These funds flow to the astronomical community via several channels:

- The Adaptive Optics Development Program (AODP)
- The Telescope System Instrumentation Program (TSIP)
- The Center for Adaptive Optics (CfAO)
- The Gemini Observatory
- NOAO and the Adaptive Optics Module for SOAR (SAM)
- The NSF PI grants program.

Figure 1 in this paper is an updated version of Figure 1 from F06. It illustrates the amount of NSF funds that have gone through the 6 channels listed above. The PI grants program is subdivided into awards given through the ATI and MRI programs and all others. For the years 1995 through 2002 (as in F06, 1995 and 1996 funds are lumped together into 1996) there are no changes to the data that went into Fig. 1 of F06 in the funding for the 6 categories. On the other hand as Table 1 shows, there are differences, some significant, for the years 2003 through 2009 between the numbers used in F06 and now, especially to the assumptions and extrapolations that were made in F06 for 2006 to 2009.[7] Table 1 has three sections: the first gives the numbers that were used for Fig. 1 in F06[8] with assumed figures in italics and smaller font; the second gives the numbers as they are now (2009, September); the third section gives the differences between the two sets of figures. Numbers in this last section that are in parenthesis indicate changes that are negative. These changes will be explained below.

---

[4] http://www.aura-astronomy.org/nv/Astro2010PanelDocs/AURAs%20assessment%20of%20AO%20V4.pdf
[5] See, for example, http://www.aura-astronomy.org/nv/Astro2010%20meeting.pdf and http://www.aura-astronomy.org/nv/Astro2010%20Gemini.pdf
[6] A description of the meeting and down-loadable presentations are at http://www.noao.edu/meetings/ao-aas/.
[7] The SOAR Adaptive Optics Module which is being constructed by NOAO out of its base budget with NSF funds was inadvertently omitted from the accounting in F06. It has been included in Table 1 with the assumption that the dollar amounts are now the same, or nearly so, as they would have been in F06.
[8] Re-examination of the grant abstracts for these years indicated that a couple of grants previously included should not have been while a couple that were *not* included, should have been. The net result of this shuffling is small. These changes were incorporated into Table 1.
3

The net change for the years 2003 to 2009 from actual and estimated numbers given in F06 to the actual numbers now is $6.4M, or a yearly average increase relative to the 2006 figures of $0.9M. Termination of the AODP has been amply compensated for by a relative increase in the grants program for 2009 and two TSIP grants in 2007. Note, though, that for 2007 to 2009 many of the differences are negative. Although the yearly increases compared to the 2006 predictions are not large, they do, at first glance appear encouraging. However, as will become apparent, this optimism may not be warranted.

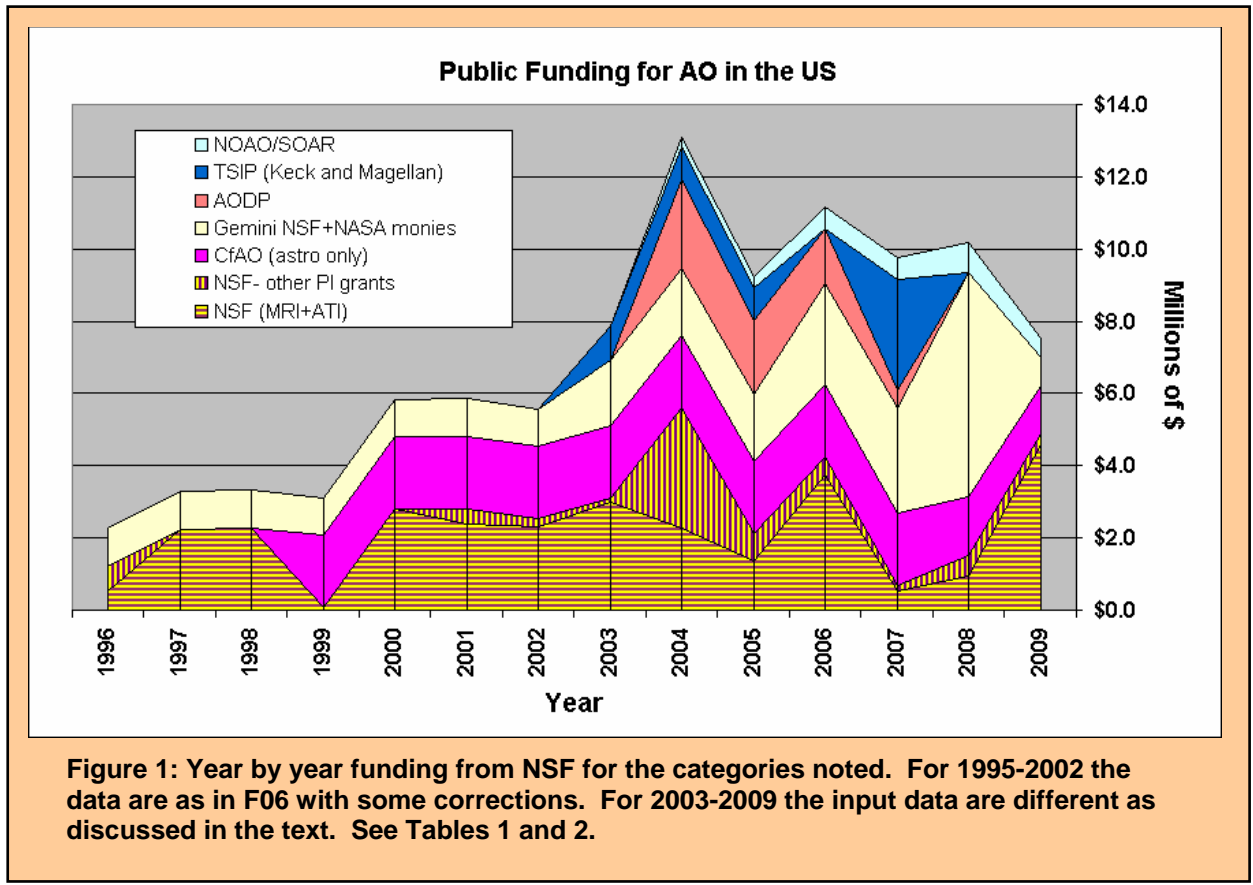

**Figure 1: Year by year funding from NSF for the categories noted. For 1995-2002 the data are as in F06 with some corrections. For 2003-2009 the input data are different as discussed in the text. See Tables 1 and 2.**

| Year | Actual and Predicted from 2006 Article in M$ | | | | | | Actual as of September 1, 2009 in M$ | | | | | | Δ (2009 – 2006) | | | | | |
|---|---|---|---|---|---|---|---|---|---|---|---|---|---|---|---|---|---|---|
| | NSF PI Grants | CfAO | AODP | TSIP | NOAO/SOAR | Gemini | NSF PI Grants | CfAO | AODP | TSIP | NOAO/SOAR | Gemini | NSF PI Grants | CfAO | AODP | TSIP | NOAO/SOAR | Gemini |
| 2003 | $2.95 | $2.00 | $0.0 | $0.91 | $0.0 | $1.85 | $3.09 | $2.00 | $0.0 | $0.91 | $0.0 | $1.85 | $0.14 | $0.0 | $0.0 | $0.0 | $0.0 | $0.0 |
| 2004 | $3.53 | $2.00 | $2.46 | $0.91 | $0.28 | $1.85 | $5.59 | $2.00 | $2.46 | $0.91 | $0.28 | $1.85 | $2.06 | $0.0 | $0.0 | $0.0 | $0.0 | $0.0 |
| 2005 | $1.13 | $2.00 | $2.05 | $0.91 | $0.29 | $1.85 | $2.12 | $2.00 | $2.05 | $0.91 | $0.29 | $1.85 | $0.99 | $0.0 | $0.0 | $0.0 | $0.0 | $0.0 |
| 2006 | $3.03 | $2.00 | $1.49 | $0.0 | $0.63 | $4.23 | $4.25 | $2.00 | $1.49 | $0.0 | $0.63 | $2.80 | $1.22 | $0.0 | $0.0 | $0.0 | $0.0 | ($1.43) |
| 2007 | *$2.0* | *$2.0* | *$2.02* | *$0.0* | *$0.63* | *$2.38* | $0.70 | $2.00 | $0.50 | $3.05 | $0.63 | $2.90 | ($1.30) | $0.0 | ($1.52) | $3.05 | $0.0 | $0.53 |
| 2008 | *$2.0* | *$2.0* | *$1.42* | *$0.0* | *$0.84* | *$2.38* | $1.53 | $1.61 | $0.0 | $0.0 | $0.84 | $6.20 | ($0.47) | ($0.39) | ($1.42) | $0.0 | $0.0 | $3.83 |
| 2009 | *$2.0* | *$0.0* | *$1.50* | *$0.0* | *$0.53* | *$2.38* | $4.87 | $1.33 | $0.0 | $0.0 | $0.53 | $0.80 | $2.87 | $1.33 | ($1.50) | $0.0 | $0.0 | ($1.58) |
| **Totals** | **$16.6** | **$12.0** | **$11.0** | **$2.72** | **$3.20** | **$16.9** | **$20.8** | **$12.9** | **$6.50** | **$5.78** | **$3.20** | **$18.3** | **$5.51** | **$0.94** | **($4.44)** | **$3.05** | **$0.0** | **$1.35** |

**Table 1**

**NSF Funding for Adaptive Optics: The View from 2006 and Now**

As I will explain below, a possibly more accurate view for 2006-2009 as well as a small extrapolation to 2010 is illustrated in Figure 2. Fig. 2 shows that the *total* amount of NSF AO funding for 2009 is about $1M more than predicted in F06 while the 2010 amount is about the

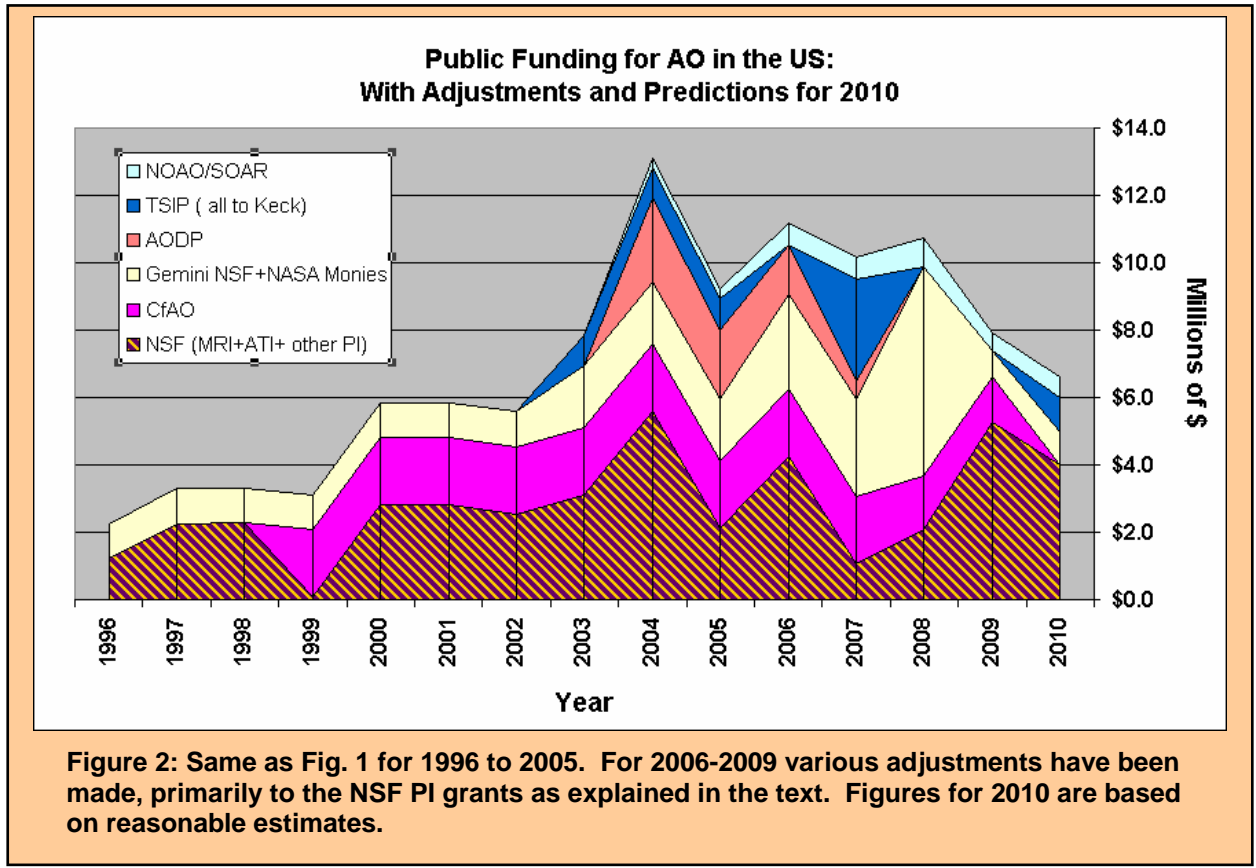

**Figure 2: Same as Fig. 1 for 1996 to 2005. For 2006-2009 various adjustments have been made, primarily to the NSF PI grants as explained in the text. Figures for 2010 are based on reasonable estimates.**

same as predicted for 2009 in F06. Overall, though, Fig. 2 shows a decline in *total* NSF AO funding from 2008 through 2010 with an ending amount of about 60% what it was during the peak years.

Next I will discuss results for each of the six channels into which NSF money for AO flows and explain the adjustments made to arrive at Fig. 2. Background information is in F06.

## The AODP

AODP was started in response to the 2000 AO Roadmap. As a result of the first call for proposals (CfP), six multi-year awards were made for a total of ~$8M with funding to start in 2004. Although there was a second CfP, no new funds were made available to fund the successful proposals. When F06 was written there was some expectation that there would be ~$1.5 million in new funds per year available for AODP starting in FY 2007. Fig. 1 of F06 assumed that this expectation would be met. It wasn't. <u>The program was effectively shut down after the first round of proposals was funded</u>. Thus, for 2010 AODP is zeroed out in Figures 1 and 2 and for years after 2007 with the exhaustion of funds from the first round of proposals.

### The CfAO

The CfAO at University of California, Santa Cruz is a NSF Science and Technology Center. CfAO has been funded through 2009 at about $4 million/year, only half of which is for astronomical AO applications. Small differences between the CfAO line here in Figs. 1 and 2 and that in F06 are due to how the funds were distributed over the past few years. The basic point, though, is that *2009 is the last year for NSF funds for the CfAO*. Therefore, CfAO is zeroed out for 2010 in Fig. 2[9].

### The TSIP

TSIP is administered by NOAO for NSF. As pointed out in F06, the only AO instrument as of 2006 that received TSIP funds was OSIRIS for Keck Observatory. For illustrative purposes its $2.75M is spread over three years in Fig. 1 here and in F06. The NOAO website[10] indicates that in 2007 the Next Generation AO system (NGAO) for Keck received $2.05M for preliminary design work and the development of an adaptive secondary for one of the Magellan telescopes received $1M. There were no new TSIP projects of any kind funded in 2008 or through August of 2009. For 2010 I assume (based on no knowledge!) that there will be one new AO project in TSIP and that it will receive $1M in funds.

### NOAO and SOAR

As part of NOAO's contribution to the operation of the SOAR telescope on Cerro Pachon, it was responsible for the design, construction, and delivering of SAM – the SOAR Adaptive Optics Module. Funds for this instrument came from NSF via NOAO's base budget. This instrument was inadvertently left out of the accounting in F06. Dollar amounts for 2004 through 2009 represent actual dollar spent (per FY). The number for 2010 is a good estimate. If needed there would be additional funds available for FY 2011.

### The Gemini Observatory

Table 1 shows some significant year to year differences between 2006 estimates and current actual values in NSF's share for funding AO activities at Gemini Observatory for 2006 to 2009. Overall, the integrated amount over this time period has increased by $1.35M. All of these changes are related to the funding stream for the Gemini Planet Imager (GPI), an "Aspen Process" instrument. For 2010 I assumed $1M, in line with current year expenditures, the conclusion of the GPI project, and the fact that there are no new approved AO projects in the instrument pipeline for Gemini[11]. One-half of the cost for other AO systems for the two Gemini telescopes (NSF's share), including MCAO and ALTAIR, are included in the yearly totals. Lacking complete information on how these expenditures were distributed from 1996 to 2003, I just put equal amounts into each year.

---

[9] It is possible that private funds will be available to keep some of the astronomical activities of the CfAO going. But my guess is that the grants program for astronomical AO R&D that CfAO administered will be curtailed or terminated.
[10] http://www.noao.edu/system/tsip/summary.php
[11] That $1M is a reasonable estimate is confirmed by Doug Simons, Gemini Director.

The NSF Grants Program.

Table 2 here is an update of Table 1 in F06.  The entire NSF grant data base for 1995 – 2009 was searched anew for the phrase "adaptive optics" or "AO" in the title, abstract, or text.  Most awards having to do with pure scientific applications of AO were easily eliminated from a reading of the titles. The resulting list was culled down to those of relevance for astronomy by reading the abstracts.   Awards whose purpose was to advance speckle imaging rather than AO were not considered.  A number of astronomy related AO awards were found in NSF's small

<div align="center">

**Table 2**

**AO Related Awards (non-Science) in NSF's ATI, MRI, and other PI Grants Programs**

</div>

| | ATI and MRI PI Grants | | Other NSF PI Grants | | All NSF PI Grants | |
|---|---|---|---|---|---|---|
| Year | # AO awards | M $ | # AO awards | M $ | Total | Δ from 2006 Summary |
| 1996/5 | 2 | $0.512 | 2 | $0.709 | $1.221 | $0.0 |
| 1997 | 3 | $2.246 | 0 | $0.0 | $2.246 | $0.0 |
| 1998 | 4 | $2.285 | 0 | $0.0 | $2.285 | $0.0 |
| 1999 | 1 | $0.071 | 0 | $0.0 | $0.071 | $0.0 |
| 2000 | 4 | $2.797 | 0 | $0.0 | $2.797 | $0.0 |
| 2001 | 3 | $2.378 | 2 | $0.430 | $2.807 | $0.0 |
| 2002 | 5 | $2.295 | 2 | $0.238 | $2.533 | $0.0 |
| 2003 | 4 | $2.989 | 1 | $0.100 | $3.089 | $0.14 |
| 2004 | 2 | $2.288 | 1 | $3.300 | $5.588 | $2.06 |
| 2005 | 2 | $1.350 | 1 | $0.766 | $2.117 | $0.16 |
| 2006 | 5 | $3.737 | 1 | $0.511 | $4.248 | $1.73 |
| 2007 | 2 | $0.547 | 1 | $0.150 | $0.697 | *($1.30)* |
| 2008 | 3 | $0.956 | 1 | $0.571 | $1.527 | *($0.47)* |
| 2009 | 5 | $4.578 | 3 | $0.290 | $4.869 | $2.87 |
| **Total** | **45** | **$29.03** | **15** | **$7.07** | **$36.10** | **$5.19** |

business program and IT research.  Topics for these awards included deformable mirrors, MEMs technology, lasers, and computer algorithms.  Specific mention would be made in the abstracts concerning the relevance of the proposed work for astronomical applications.  But most of the relevant awards were given through the ATI program of NSF's AST division and, a smaller number, through the NSF-wide MRI program, many of which are jointly funded with the ATI program.  There were also a couple of awards in NSF's ATM division for AO development work related to solar astronomy.

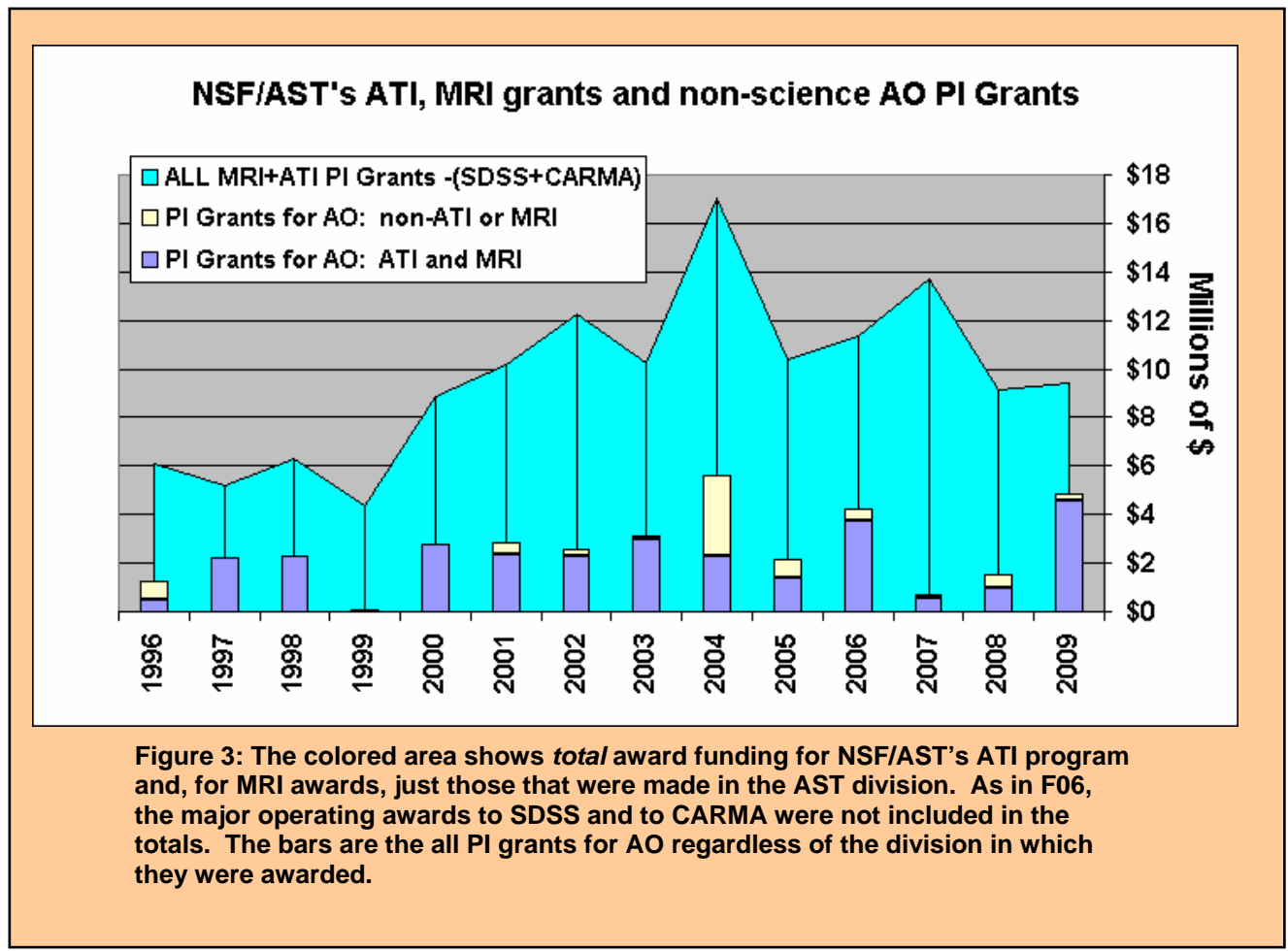

**Figure 3: The colored area shows *total* award funding for NSF/AST's ATI program and, for MRI awards, just those that were made in the AST division.  As in F06, the major operating awards to SDSS and to CARMA were not included in the totals.  The bars are the all PI grants for AO regardless of the division in which they were awarded.**

Table 2 distinguishes awards made through the ATI and MRI programs from all other AO related awards.  The two rightmost columns give the totals and the differences from the totals given in Table 1 of F06.  Some of the positive differences between 2003 and 2006 relate to the difference between a "standard" and a "continuing" grant.  For a *standard* grant the total funds awarded are specified at the outset and given in NSF's online database.  For *continuing* grants the amount awarded as given in the on-line database refers just to the initial amount or to the total amount awarded to date.  More funds can be given to the grant as time passes.  Thus, in the three years since I completed the work for F06, continuing awards with a start date in 2004-2006 continued to receive funds.  In addition to half of the 2004 to 2006 awards being continuing grants ,three of the seven awards with starting date between 2007 and 2009 are as well as and thus have received considerable additional funds since their start dates.

Two further comments on Table 2 and Fig. 1: The big up-tick for the 2004 awards is due entirely to the collaborative effort led by Gemini and including the USAF-SOR and Keck to obtain high power lasers for astronomical applications. Second, the increase for 2006 is attributable to one new award issued after the F06 article was completed as well as substantial amounts of new funds added to the three continuing grants first awarded in 2006. The negative entries for the deltas for 207 and 2008 may be attributable in part to the continuing grants not having received all of their funding

The award total for 2009 is substantially higher than the F06 prediction. In fact, if you exclude the large grant to Gemini for laser development in 2004, the 2009 award total is greater than that of any other year in Table 2. Based on a reading of the award abstracts (some of which note explicitly that they are from monies from the ARRA, a.k.a. "stimulus money") for 2009 and conversations with individuals at NSF, a substantial fraction of the 2009 AO award monies can be attributable to the one-off stimulus funding. In addition, stimulus funds to other areas of astronomy within NSF may have freed up additional funds for AO specific grants.

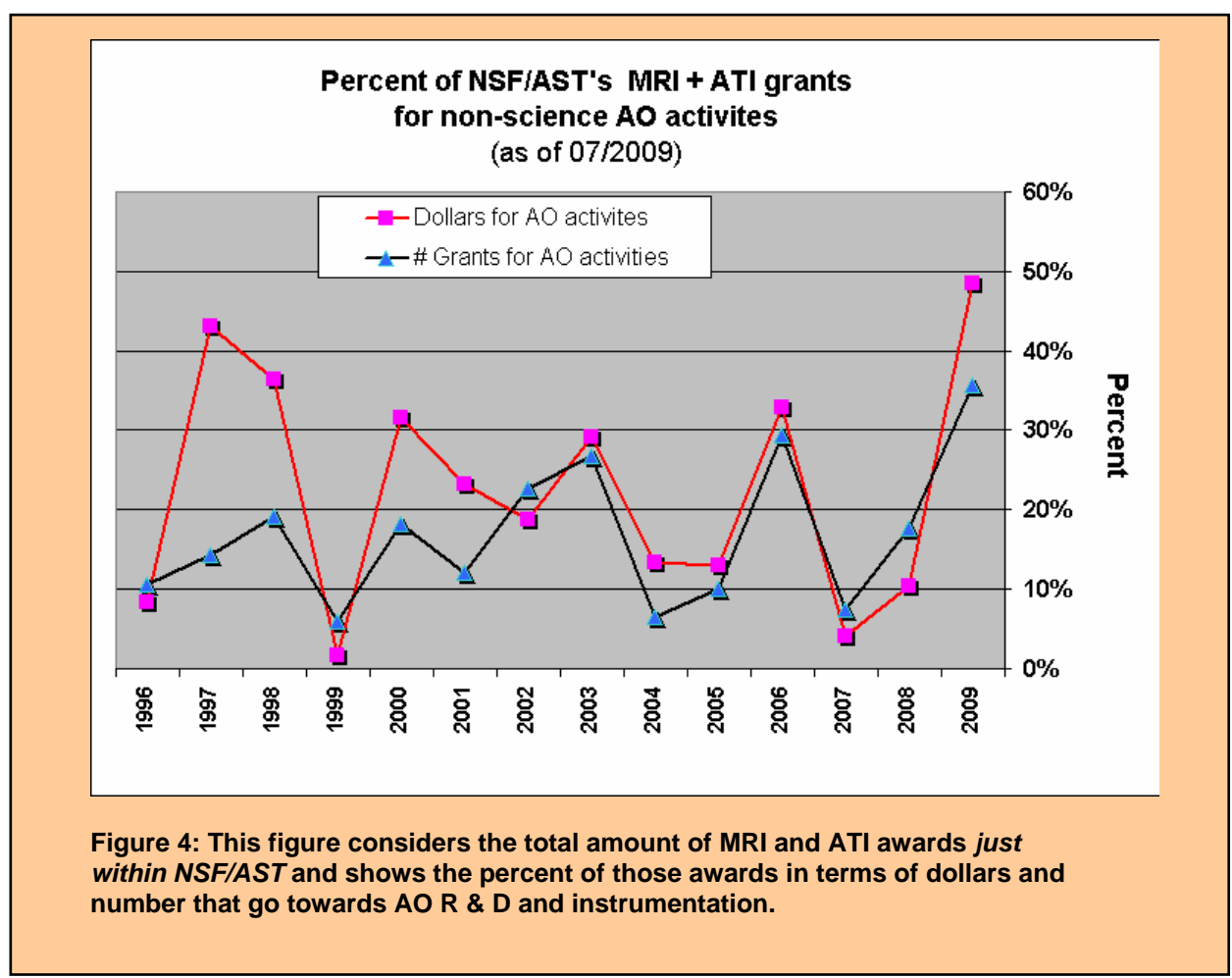

**Figure 4: This figure considers the total amount of MRI and ATI awards *just within NSF/AST* and shows the percent of those awards in terms of dollars and number that go towards AO R & D and instrumentation.**

So to construct Figure 2 I did the following: First, to correct for additional funds that might flow into continuing grants I took the 2004 through 2006 continuing grants and calculated that on average they had increased between by a factor of 1.76 between 2006 and now. I applied this correction factor to the 2007 to 2009 continuing grants and used these adjusted numbers – and the unchanged standard grant numbers – for Figure 2. As may be seen from a comparison of Figs. 1 and 2, the fraction change to the total NSF PI awards monies for each year is modest. Based on the adjusted numbers for 2009 and that an unknown fraction of the new money for 2009 can be attributed to one-off stimulus funding, I assumed that the eventual amount that would be granted to AO awards made during the year 2010 would be $1M less that the adjusted 2009 awards.

Figure 3, similar to Fig. 2 in F06, is based on the data in Table 2 and the total amount of NSF/AST ATI funding plus the total amount of MRI awards made just in the AST division. Since essentially all AO awards come out of the ATI and MRI pots, this figure shows the total potential amount available for AO related work. Figure 4 shows, for the MRI and ATI awards, the fractions of the awarded funds and the number of awards, that went for AO related work. Fig. 3 shows that as of September 1, 2009 the total for all ATI and all AST/MRI grants is slightly above the total for 2008 but still $1 to $2M below the peak years for *all* astronomy related ATI and MRI grants, 2002 through 2007, even after subtracting the 2004 Gemini Laser grant. Figure 4 shows that in 2009 AO related awards received a higher fraction of total ATI and AST/MRI funds and awards than at any time in the past 15 years.

Aside from AO funding within the AST and ATM divisions of NSF, since 2005, $1.9M has been awarded for AO R&D via NSF in the Small business Innovation Research (SBIR) program. This is 15% of the total NSF funds awarded for AO research over the same time period. A recent article on the SBIR program appeared in *Science*[12].

### Table 3

Institutions Receiving the Most Funds From NSF for AO Related Work (non-Science) 1995 - 2009

| Institution | # AO awards | M $ | % of Total $$ |
|---|---|---|---|
| Univ. Arizona | 8 | $6.66 | 18% |
| Univ. Hawai'i | 6 | $5.97 | 17% |
| CARA | 2 | $3.31 | 9% |
| Gemini | 1 | $3.30 | 9% |
| NJIT | 4 | $2.93 | 8% |
| Caltech | 6 | $2.64 | 7% |
| UCSC | 2 | $2.20 | 6% |
| **Totals** | **29** | **$27.0** | **75%** |

As noted in F06, the NSF PI awards for AO related non-science work are heavily concentrated in terms of both the individuals who are the PIs and the institutions which receive the awards and that the linkage between the two groups is tight. Table 3 lists the top 7 institutions in terms of awarded funds in the 15 years between 1995 and 2009. They were selected in order of funds received for AO work until I crossed the 70% mark of total funds awarded. In F06 it took only 5 institutions to cross the 70% line. The newcomers are Caltech, Gemini, and UCSC, which received substantial new awards and/or additions to existing continuing grants. Four of these 7 were on the earlier list in the F06. The Universities of Hawai'i and Arizona continue to dominate the rankings. The AMNH is no longer on the list because I had mistakenly put a grant for speckle

---

[12] *Science*, 2009, vol. **325**, p. 18.



work into the AO category. The dominance of these 7 institutions in terms of AO funding is underscored by the fact that after UCSC and Caltech, the next three largest amounts to single institutions are between $1.0 and $1.4M. Thus, only 7 institutions, one-quarter of the total number of distinct institutions receiving AO related awards, have received nearly half of the total number of awards and three-quarters of the total funds.

For Table 4 I selected the PIs in order of funds received for AO work over the 15 year time period until I crossed 70% of the total. This list has the same scientists as Table 3 of F06 with the addition of R. Dekany, D. Gavel, R. Angel, and Ed Kibblewhite. So 13 individuals out of 45 distinct PIs on AO related non-science grants, account for just over 70% of the funds awarded and half of the number of awards. Except for Kibblewhite, Thompson, and Oppenheimer from the Universities of Chicago and Illinois and from the AMNH, respectively, the home institutions of the individuals in Table 4 correspond to the institutions that collectively have received the largest sums of grant money.

| Table 4 | | | | |
|---|---|---|---|---|
| Individual PIs Receiving the Most Funds From NSF for AO Work (non-Science) 1995 - 2009 | | | | |
| PI | Inst. | #AO Awards | M $ | % of Total $$ |
| Wizinowich | CARA | 2 | $3.31 | 9% |
| Simons | Gemini /AURA | 1 | $3.30 | 9% |
| Ftaclas | U. Hawaii | 3 | $3.06 | 8% |
| Lloyd-Hart | U. Arizona | 3 | $2.78 | 8% |
| Gavel | UCSC | 2 | $2.20 | 6% |
| Lin | U. Hawai'i | 1 | $1.98 | 5% |
| Rimmele | NJIT | 1 | $1.82 | 5% |
| Dekany | Caltech | 3 | $1.56 | 4% |
| Close | U. Arizona | 1 | $1.37 | 4% |
| Kibblewhite | U. Chicago | 2 | $1.25 | 3% |
| Thompson | U. Illinois | 1 | $1.23 | 3% |
| Oppenheimer | AMNH | 3 | $1.03 | 3% |
| Angel | U. Arizona | 1 | $0.95 | 3% |
| **Totals** | **13** | **24** | **$25.8** | **72%** |

## **Summary and Conclusions**

> *"Current projections indicate that AO implementation on public and private telescopes in the U.S. will soon seriously lag that on the ESO VLT as measured by funds available. There needs to be a significant infusion of public funds for AO development (through AODP) and for AO implementation (through TSIP) so that when combined with private funds, the U.S. astronomical community as a whole can take full advantage of AO systems on both public and private telescopes."*

The key conclusions of F06 are summarized in the above quote. The purpose of the present article was to test the validity of the projections made in F06 and to examine the state of public funding for AO R&D and instrumentation in the US. I have not re-investigated the current state of private funding, nor of the funding for AO activities at ESO, Subaru, etc. The conclusions are:

- The F06 projection for *public* AO funding in the US for 2009 is about $1M less than the actual value. There also has been considerable shuffling amongst the various channels for the distribution of these funds. Some reasonable assumptions for funding in 2010 point to a continuing decline in the amount of public funding for AO work.

- In F06 I assumed that the AODP funding line would be brought back with about $1.5M per year in new funds. This has not happened. In the course of several public forums this past year NSF spokespeople have indicated that the gap left by AODP's demise would be filled by ATI grants. Indeed, an NSF posting for the ATI program dated Dec 12, 2008 states[13]: "*Proposals in the area of adaptive optics that address the community-developed 2008 Adaptive Optics Roadmap for future development … should make clear the relation of the proposed work to the Roadmap goals*." This approach may well account for the significant uptick in AO funding via the MRI and ATI lines. But I point out that nearly all of the new 2009 funds are for specific instruments and system development rather than for fundamental R&D in AO as was the stated purpose of the AODP. Also, probably a significant fraction of these new funds are a result of the stimulus package is a "one-time" event.

- NSF's support for astronomically related AO activities at the CfAO terminates this year as expected. This removes $2M per year from the AO funding line.

- NSF support for AO at SOAR via NOAO is a one off deal for the construction of just one instrument - SAM

- The Telescope System Instrumentation Program (TSIP) is a possible bright spot – it provided $3M in funding in 2007. Although as of this writing no new instruments of any kind have been funded for 2008 or 2009, it is expected that new funds will be available later this year or early in 2010; but whether or not any of these funds will be for an AO instrument remains to be seen.

---

[13] http://www.nsf.gov/funding/pgm_summ.jsp?pims_id=5660&org=AST

14- In F06 I assumed that NSF's share of funding AO activities at the Gemini Observatory would remain at around $2M per year. While there was a sharp peak in 2008 corresponding to a major uptick in construction of GPI, for 2009 funding from NSF declined to less than $1M (Fig. 1 and 2).

The net result of the actual funding streams for 2007 to 2009 with a small extrapolation to 2010 shows:

- Actual public funding for AO in 2007 and 2008 was about $3M higher per year than predicted in F06. This was entirely due to new TSIP instruments and spending for GPI at the Gemini Observatory. These offset the termination of AODP funding.

- Total public funding for AO in the US has declined to $7M per year for 2009; less than that is predicted $6M for 2010.

The bottom line is that public, i.e. NSF, funding for AO in the US in 2010 will be at its lowest level since the 2000 to 2002 period. If private funds in the US are at the levels predicted in F06 and if there has been no change in the funding level at ESO for AO, then ESO is outspending the US by at least a factor of three. And, to repeat, ESO funds are concentrated on one observatory. Those in the US are spread over about one dozen institutions.

I thank Doug Simons (Gemini) Bob Blum and Dave Sprayberry (NOAO), and Jeff Pier (NSF) for some clarifying information and remarks.

14